\documentclass[conference]{IEEEtran}
\IEEEoverridecommandlockouts

\usepackage{orcidlink}
\usepackage{cite}
\usepackage{amsmath,amssymb,amsfonts}
\usepackage{bm}
\usepackage{algorithmic}
\usepackage{graphicx}
\usepackage{textcomp}
\usepackage{xcolor}
\usepackage{dsfont}

\usepackage[nolist, nohyperlinks]{acronym}
\usepackage{subcaption}
\usepackage{xcolor}
\usepackage{comment}
\usepackage{cite}
\usepackage{soul}

\usepackage{epstopdf}
\usepackage{xfrac}
\usepackage{pifont}
\newcommand{\cmark}{\textcolor{green}{\ding{51}}}%
\newcommand{\xmark}{\textcolor{red}{\ding{55}}}%

\usepackage{csquotes}
\usepackage{booktabs}
\usepackage{multirow}
\usepackage{tabularx,ragged2e}
\newcolumntype{Y}{>{\centering\arraybackslash}X}
\usepackage{footmisc}
\usepackage{hyperref}
\usepackage{cleveref}

\usepackage{svg}
\usepackage{varwidth}
\usepackage{tikz}
\usepackage{pgf}
\usepackage{pgfplots}
\usepackage{pgfplotstable}
\usetikzlibrary{positioning,backgrounds,fit,calc,patterns,arrows.meta}
\pgfplotsset{compat=1.3}
\usetikzlibrary{shapes.geometric}
\usetikzlibrary{shapes.arrows}
\usetikzlibrary{arrows.meta}
\usepackage[skins]{tcolorbox}
\usepackage{adjustbox}
\usepackage{array}

\definecolor{lightgray}{gray}{0.5}

\title{Investigating Training Objectives for\\ Generative Speech Enhancement
\thanks{We acknowledge the support by the German Research Foundation (DFG) in the transregio project Crossmodal Learning (TRR 169).}
}

\author{
\IEEEauthorblockN{Julius Richter \orcidlink{0000-0002-7870-4839}}
\IEEEauthorblockA{\textit{Signal Processing Group} \\
\textit{University of Hamburg}\\
Hamburg, Germany}
\and
\IEEEauthorblockN{Danilo de Oliveira \orcidlink{XXX}}
\IEEEauthorblockA{\textit{Signal Processing Group} \\
\textit{University of Hamburg}\\
Hamburg, Germany}
\and
\IEEEauthorblockN{Timo Gerkmann \orcidlink{0000-0002-8678-4699}}
\IEEEauthorblockA{\textit{Signal Processing Group} \\
\textit{University of Hamburg}\\
Hamburg, Germany 
}
}

\begin{document}

\begin{acronym}
\acro{sgm}[SGM]{score-based generative model}
\acro{snr}[SNR]{signal-to-noise ratio}
\acro{gan}[GAN]{generative adversarial network}
\acro{vae}[VAE]{variational autoencoder}
\acro{ddpm}[DDPM]{denoising diffusion probabilistic model}
\acro{STFT}[STFT]{short-time Fourier transform}
\acro{iSTFT}[iSTFT]{inverse short-time Fourier transform}
\acro{SDE}[SDE]{stochastic differential equation}
\acro{ODE}[ODE]{ordinary differential equation}
\acro{ou}[OU]{Ornstein-Uhlenbeck}
\acro{VE}[VE]{variance exploding}
\acro{OUVE}[OUVE]{Ornstein-Uhlenbeck process with variance exploding}
\acro{dnn}[DNN]{deep neural network}
\acro{PESQ}[PESQ]{Perceptual Evaluation of Speech Quality}
\acro{se}[SE]{speech enhancement}
\acro{tf}[T-F]{time-frequency}
\acro{elbo}[ELBO]{evidence lower bound}
\acro{WPE}{weighted prediction error}
\acro{MAC}{multiply–accumulate operation}
\acro{PSD}{power spectral density}
\acro{RIR}{room impulse response}
\acro{SNR}{signal-to-noise ratio}
\acro{LSTM}{long short-term memory}
\acro{POLQA}{Perceptual Objectve Listening Quality Analysis}
\acro{SDR}{signal-to-distortion ratio}
\acro{SI-SDR}{scale invariant signal-to-distortion ratio}
\acro{ESTOI}{Extended Short-Term Objective Intelligibility}
\acro{ELR}{early-to-late reverberation ratio}
\acro{TCN}{temporal convolutional network}
\acro{DRR}{direct-to-reverberant ratio}
\acro{nfe}[NFE]{number of function evaluations}
\acro{rtf}[RTF]{real-time factor}
\acro{MOS}[MOS]{mean opinion scores}
\acro{EMA}[EMA]{exponential moving average}
\acro{SB}[SB]{Schrö\-din\-ger bridge}
\acro{SGMSE}[SGMSE]{score-based generative models for speech enhancement}
\acro{EDM}[EDM]{elucidating the design space of diffusion-based generative models}
\acro{GPU}[GPU]{graphics processing unit}
\acro{VB-DMD}[VB-DMD]{Voiceband-Demand}
\acro{SB-VE}[SB-VE]{Schrödinger bridge with variance exploding diffusion coefficient}
\end{acronym}

%
\maketitle
\begin{abstract}
Generative speech enhancement has recently shown promising advancements in improving speech quality in noisy environments. Multiple diffusion-based frameworks exist, each employing distinct training objectives and learning techniques. This paper aims to explain the differences between these frameworks by focusing our investigation on score-based generative models and the Schrödinger bridge. We conduct a series of comprehensive experiments to compare their performance and highlight differing training behaviors. Furthermore, we propose a novel perceptual loss function tailored for the Schrödinger bridge framework, demonstrating enhanced performance and improved perceptual quality of the enhanced speech signals. All experimental code and pre-trained models are publicly available to facilitate further research and development in this domain\footnote{\url{https://github.com/sp-uhh/sgmse}}.
\end{abstract}

\section{Introduction}
\label{sec:intro}

Diffusion-based generative models have been successfully employed in various audio restoration tasks, most notably in speech enhancement~\cite{lemercier2025diffusion}. Generative methods in this task aim at estimating and sampling from the clean speech distribution conditioned on noisy speech. Unlike predictive models, generative models enable the generation of multiple valid estimates for a given input and can be utilized for generalized (or universal) speech enhancement, effectively addressing various corruption types~\cite{richter2024causal}.

Numerous diffusion-based generative approaches exist, all centered around the idea of defining a transformation between the data distribution and a tractable prior distribution (e.g., Gaussian). Popular frameworks include continuous-time diffusion models~\cite{song2021sde}, EDM~\cite{karras2022elucidating}, flow matching~\cite{lipman2023flow}, and the \ac{SB}~\cite{chen2021likelihood}. Each of these approaches has been applied to speech enhancement.

\Ac{SGMSE}~\cite{welker2022speech, richter2023speech} employs continuous-time diffusion models based on \acp{SDE}. Follow-up work utilizes the EDM framework; it has been proposed to use a change of variable to consider the \ac{SDE} satisfied by the environmental noise~\cite{gonzalez2024investigating}, which results in the required linear affine drift term. Flow matching has been used in SpeechFlow~\cite{liu2024generative}, where the authors apply masked audio prediction as a self-supervised pretraining technique.

More recently, the \ac{SB} has been proposed for speech enhancement~\cite{jukic2024schr}. The \ac{SB} is a generative model that seeks an optimal way to transport one probability distribution to another distribution~\cite{chen2021likelihood}. This approach enables starting the generative process directly from the noisy input and allows for using a data prediction loss \cite{chen_2023_sb}. 

This paper builds upon the above-mentioned advances and explores multiple training objectives and learning techniques for generative speech enhancement. We begin by exploring score-based generative models and connecting various loss functions used to learn the score function. For further details, we refer to \cite{kingma2023understanding}, showing how various diffusion-based generative model objectives can be understood as special cases of a weighted loss.  Second, we examine the \ac{SB} approach for speech enhancement~\cite{jukic2024schr} and establish a connection to \ac{SGMSE}~\cite{richter2023speech}. 
Moreover, we propose a novel perceptual loss function for the \ac{SB} framework and perform ablation studies to evaluate its effect.
Perceptually motivated loss functions have been extensively integrated into speech enhancement frameworks~\cite{koizumi2018dnn, fu2019metricgan}. However, their use in diffusion-based generative models, particularly within the Schrödinger bridge framework, is a novel approach. We specifically choose the PESQ metric because it is a widely used instrumental measure in speech enhancement for assessing speech quality and, unlike POLQA~\cite{polqa2018}, a differentiable version is available~\cite{kim2019end}.

Our experiments demonstrate that score-based generative models trained with different objective functions exhibit varying training behaviors, although they theoretically model the same underlying concepts. We hypothesize this is due to the neural network's different training tasks. Additionally, we show that our novel perceptual loss for the \ac{SB} achieves state-of-the-art performance in PESQ on the \ac{VB-DMD} benchmark~\cite{valentini2016investigating}.

Contemporaneously to our work, Wang et al. \cite{wang2024diffusion} explore the \ac{SB} and set a symmetric noise scheduling, where the diffusion shrinks at both boundaries. Furthermore, they combine the SB concept with a two-stage approach inspired by StoRM~\cite{lemercier2023storm} by aiding the generative model with a magnitude ratio mask.

The paper is structured as follows: Sec. \ref{sec:methods} explores \ac{SGMSE} and the \ac{SB} framework. This is followed by describing the experimental setup in Sec. \ref{sec:experiments}. Then, we present the results of the experiments in Sec. \ref{sec:results}. Finally, in Sec. \ref{sec:conclusion}, we summarize our key findings and provide an outlook on future work.

\section{Methods}
\label{sec:methods}

This section discusses two existing generative approaches for speech enhancement and explores their connection. First, we introduce \ac{SGMSE}~\cite{richter2023speech}. Second, we examine the \ac{SB} for speech enhancement~\cite{jukic2024schr}. Both approaches are diffusion-based stochastic processes aiming to model and manipulate probability distributions. Score-based models emphasize learning score functions, whereas the SB approach can be considered as an optimal transport problem.

\subsection{Score-based generative models for speech enhancement}

Following \cite{richter2023speech}, the diffusion forward process is described by the solution to the \emph{forward \ac{SDE}}
\begin{equation} \label{eq:ouve-sde}
    \mathrm{d} \mathbf{x}_t = \gamma (\mathbf{y} - \mathbf{x}_t) \mathrm{d} t + g(t) \mathrm{d} \mathbf{w},
\end{equation}
where $\mathbf{x}_t\!\in\!\mathbb{C}^d$ is the process state at time $t \!\in\! \left[0, 1 \right]$ and $\mathbf y\!\in\!\mathbb{C}^d$ is the noisy speech. The diffusion coefficient $g(t)=\sqrt{c} k^t$ controls the Gaussian noise introduced by the Wiener process $\mathbf w\!\in\!\mathbb{C}^d$. Moreover, $\gamma$, $c$, and $k$ are positive scalar constants that are set as hyperparameters. The forward process is also called \ac{OUVE} \ac{SDE}\cite{lay23_interspeech}. 

The marginals of the time-reversed forward process can be represented as marginals of another stochastic process (see Theorem A.1 in \cite{berner2024optimal}). This resulting process is described by the solution to the so-called \emph{reverse \ac{SDE}}
\begin{equation}
  \label{eq:ouve_reverse}
  \mathrm d \mathbf x_t =  [ -\gamma(\mathbf y - \mathbf x_t)
  + g(t)^2 \nabla_{\mathbf x_t} \log p_t(\mathbf x_t | \mathbf y)
  ] \mathrm d t + g(t) \mathrm d \bar{\mathbf w}\,,
\end{equation} 
where $\nabla_{\mathbf x_t}\log p_t(\mathbf x_t | \mathbf y)$ is the conditional score function, and $\bar{\mathbf w}$ is the Wiener process backward in time.

It can be shown that the \ac{OUVE} \ac{SDE} results in an interpolation between clean speech $\mathbf x_0$ and noisy speech $\mathbf y$ with exponentially increasing variance~\cite{richter2023speech}. The evolution of the marginals is described by the time-dependent mean
\begin{equation}
    \boldsymbol\mu_t(\mathbf x_0, \mathbf y) = e^{-\gamma t} \mathbf x_0 + (1 - e^{-\gamma t} ) \mathbf y
\end{equation}
and the time-dependent variance
\begin{equation}
    \sigma_t^2 = \frac{c(k^{2t}-e^{-2\gamma t})}{2(\gamma + \log k)}\,,
\end{equation}
that allow for direct sampling of the process state $\mathbf x_t$ at time $t$ using the perturbation kernel 
\begin{equation}
\label{eq:perturbation-kernel}
    p_{0t}(\mathbf x_t|\mathbf x_0, \mathbf y) = \mathcal{N}_\mathbb{C}(\mathbf x_t; \boldsymbol\mu_t(\mathbf x_0, \mathbf y), \sigma_t^2 \mathbf{I})\,.
\end{equation}

The score function is typically intractable and approximated by a score model $\mathbf s_\theta$ with parameters $\theta$. To train the score model, we use the denoising score-matching objective~\cite{vincent2011connection}
\begin{equation}\label{eq:score-matching}
    \mathcal L_\text{DSM} = \lambda(t)
        \lVert \mathbf s_\theta(\mathbf x_t, \mathbf y, t) -  \nabla_{\mathbf x_t} \log p_{0t}(\mathbf x_t|\mathbf x_0, \mathbf y)\rVert_2^2
\end{equation}
where $\lambda(t)$ is a weighting function, and the other variables are sampled according to $t \sim \mathcal U[0,1]$, $(\mathbf x_0, \mathbf y)\sim p_\text{data}(\mathbf x_0, \mathbf y)$ from the dataset, and $\mathbf x_t \sim p_{0t}(\mathbf x_t|\mathbf x_0, \mathbf y)$. The score matching loss is essentially equivalent to a noise prediction loss
\begin{equation}
    \label{eq:loss_score}
    \mathcal L_\text{score} = 
        \lVert  \mathbf s_\theta(\mathbf x_t, \mathbf y, t) \,\sigma_t +  \mathbf z \rVert_2^2 ~\text{,}
\end{equation}
when $\lambda(t) = \sigma_t^2$ in Eq.~\eqref{eq:score-matching}, and $\mathbf z \sim \mathcal N(\mathbf 0, \mathbf I)$. To improve numerical stability, the output of an employed neural network $F_\theta$ is often scaled by a factor of  $1/\sigma_t$ such that $\mathbf s_\theta(\mathbf x_t, \mathbf y, t) = F_\theta(\mathbf x_t, \mathbf y, t) / \sigma_t$.

Following the derivations in \cite{karras2022elucidating}, it can also be shown that denoising score matching for \ac{SGMSE} is equivalent to training a denoiser model $D_\theta$ with the denoising loss
\begin{equation}
\label{eq:denoising_loss}
 \mathcal{L}_\text{denoise} = \lambda(t) \lVert D_\theta(\mathbf x_t, \mathbf{y}, t) - \boldsymbol{\mu}_t(\mathbf x_0, \mathbf y) \rVert^2_2\,. 
\end{equation}
Furthermore, it was argued that it is beneficial to precondition a neural network $F_\theta$ to obtain the denoiser
\begin{equation}
\label{eq:preconditioning}
    D_\theta (\mathbf x_t, \mathbf y, t) =c_\text{skip}(t) \mathbf x_t + c_\text{out}(t) F_\theta (c_\text{in}(t) \mathbf x_t,\, c_\text{in}(t) \mathbf y,\, t)
\end{equation}
where $c_\text{skip}(t)$, $c_\text{out}(t)$, $c_\text{in}(t)$, and $c_\text{in}(t)$ are time-dependent functions that can be derived from first principles (see Appendix B.6 in \cite{karras2022elucidating}). Then, the score is then given by 
\begin{equation}
    \mathbf s_\theta(\mathbf x_t, \mathbf y, t) = \frac{D_\theta(\mathbf x_t, \mathbf y, t) - \mathbf x_t}{\sigma_t^2}\,.
    \label{eq:connection_score_denoiser}
\end{equation}

\subsection{Schrödinger bridge for speech enhancement}

The \ac{SB} \cite{schrodinger1932theorie} is defined as the minimization of the Kullback-Leibler divergence $D_\text{KL}$ between a path measure $p$ and a reference path measure $p_\text{ref}$, subject to boundary conditions
\begin{equation}
  \label{eq:sb}
  \min_{p \in \mathcal{P}_{\left[0, 1\right]}} D_\text{KL} \left(p, p_\mathrm{ref} \right)
  \quad
  \text{s. t.}
  \quad
  p_0 = p_x ,
  \thickspace
  p_1 = p_y ,
\end{equation}
where $\mathcal{P}_{\left[0, 1\right]}$ is the space of path measures on $\left[0, 1\right]$~\cite{chen2021likelihood}.  An optimal transport solution is given by a pair of symmetric forward and reverse SDEs, with the forward \ac{SDE} being
\begin{equation}
  \label{eq:sb_forward_sde}
  \mathrm d \mathbf{x}_t  = \left[ \mathbf{f}(\mathbf x_t) + g(t)^2 \nabla_{\mathbf x_t} \log \Psi_t(\mathbf{x}_t) \right] \mathrm d t + g(t) \mathrm d \mathbf{w}_t, \, \mathbf{x}_0 \sim p_x , 
\end{equation}
and the reverse SDE being
\begin{equation}
  \label{eq:sb_reverse_sde}
  \mathrm d\mathbf{x}_t  = \left[ \mathbf{f}(\mathbf x_t) - g(t)^2 \nabla_{\mathbf x_t}\log \bar{\Psi}_t(\mathbf{x}_t) \right] \mathrm d t + g(t) \mathrm d \bar{\mathbf{w}}_t, \, \mathbf{x}_1 \sim p_y ,
\end{equation}
where the functions $\Psi_t, \bar{\Psi}_t$ are described by coupled partial differential equations (see Theorem 1 in \cite{chen2021likelihood})
\begin{align}
\label{eq:SB-PDE}
\begin{cases}
    \frac{\partial \Psi_t}{\partial t}=-\nabla_{\mathbf{x}_t} \Psi_t(\mathbf{x}_t)^\text{T} \mathbf{f}(\mathbf{x}_t)-\frac{1}{2} \operatorname{Tr}(g(t)^2 \nabla_{\mathbf{x}_t}^2 \Psi_t(\mathbf{x}_t)) \\
    \frac{\partial \bar{\Psi}_t}{\partial t}=-\nabla_{\mathbf{x}_t} \cdot(\bar{\Psi}_t(\mathbf{x}_t) \mathbf{f}(\mathbf{x}_t))+\frac{1}{2} \operatorname{Tr}(g(t)^2 \nabla_{\mathbf{x}_t}^2 \bar{\Psi}_t(\mathbf{x}_t)) \notag
\end{cases} \\
\textit{s.t.}\,\, \Psi_0\bar\Psi_0 = p_x, \Psi_1\bar\Psi_1 = p_y.
\end{align}

However, for a system of symmetric forward and reverse SDEs in Eqs.~(\ref{eq:sb_forward_sde}, \ref{eq:sb_reverse_sde}), and arbitrary $\Psi_t$ and $\bar{\Psi}_t$, there are infinitely many solutions bridging the prior to the target~\cite{richter2024improved}. According to Nelson's identity \cite{nelson1967dynamical}
\begin{equation}
     \nabla_{\mathbf x_t} \log \Psi_t(\mathbf x_t)+ \nabla_{\mathbf x_t} \log \bar{\Psi}_t(\mathbf x_t) = \nabla_\mathbf{x_t} \log p_t(\mathbf x_t) ,
\end{equation}
which is a necessary condition for time-reversal \cite{richter2024improved}, we note that in score-based generative models, this corresponds to setting $\nabla_{\mathbf x_t} \log \Psi_t$ to zero. This implies that in score-based generative models, the drift of the forward process, $\mathbf f_\text{SGMSE}(\mathbf x_t) = \gamma (\mathbf{y} - \mathbf{x}_t)$ in Eq.~\eqref{eq:ouve-sde}, has to be chosen such that the perturbation kernel is known analytically. However, this is not required for the general \ac{SB} formulation.

Although solving the general \ac{SB} is typically intractable, closed-form solutions are available for specific cases, such as those involving Gaussian boundary conditions~\cite{bunne2023schrodinger}. Assume a drift $\mathbf{f}(\mathbf{x}_t)\!=\!f(t) \,\mathbf{x}_t$ and conditional Gaussian boundary conditions $p_0(\mathbf x | \mathbf x_0)\!=\!\mathcal{N}_\mathbb{C} \left(\mathbf x; \mathbf{x}_0, \epsilon_0^2 \mathbf{I} \right)$ and $p_1(\mathbf x | \mathbf y)\!=\!\mathcal{N}_\mathbb{C} \left(\mathbf x;  \mathbf{y}, \epsilon_1^2 \mathbf{I} \right)$ where $\epsilon_1 = \mathrm{e}^{\int_0^1 f(\tau) \mathrm d {\tau}} \epsilon_0$. For $\epsilon_0\!\to\!0$, the SB solution between clean speech $\mathbf{x}_0$ and noisy speech $\mathbf{y}$ can be expressed as
\begin{equation}
  \label{eq:sb_solution}
  \bar{\Psi}_t(\mathbf x_t | \mathbf x_0) = \mathcal{N}_\mathbb{C}( \alpha_t \mathbf{x}_0, \alpha_t^2 \sigma_t^2 \mathbf{I}),
  \;
  \Psi_t(\mathbf x_t | \mathbf y) = \mathcal{N}_\mathbb{C}( \bar{\alpha}_t \mathbf{y}, \alpha_t^2 \bar{\sigma}_t^2 \mathbf{I}) 
\end{equation}
with parameters $\alpha_t\!=\!\mathrm{e}^{\int_0^t f(\tau) \mathrm d {\tau}}$, $\sigma_t^2\!=\!\int_0^t \frac{g^2(\tau)}{\alpha_\tau^2} \mathrm d {\tau}$, $\bar{\alpha}_t\!=\!\alpha_t \alpha_1^{-1}$ and $\bar{\sigma}_t^2 = \sigma_1^2 - \sigma_t^2$ \cite{chen_2023_sb}. Therefore, the marginal distribution is the Gaussian distribution 
\begin{equation}
  \label{eq:sb_marginal_distribution}
  p_t(\mathbf x_t | \mathbf x_0, \mathbf y) = \mathcal{N}_\mathbb{C} \left(\mathbf x_t; \bm{\mu}_t(\mathbf x_0, \mathbf y), \sigma_{\mathbf x_t}^2 \mathbf{I} \right)
\end{equation}
whose mean and variance are defined as
\begin{equation}
  \label{eq:sb_mean_variance}
  \bm{\mu}_t(\mathbf x_0, \mathbf y) = w_x(t) \mathbf{x}_0 + w_y(t) \mathbf{y} ,
  \quad
  \sigma_{\mathbf x_t}^2 = \frac{ \alpha_t^2 \bar{\sigma}_t^2 \sigma_t^2 }{ \sigma_1^2 } ,
\end{equation}
with $w_x(t) = \alpha_t \bar{\sigma}_t^2 / \sigma_1^2$, and $w_y(t) = \bar{\alpha}_t \sigma_t^2 / \sigma_1^2$~\cite{chen_2023_sb}.

In this paper, we adopt the same \ac{VE} diffusion coefficient $g(t)= \sqrt{c}k^t$ as used in Eq. \eqref{eq:ouve-sde}, and set $f(t)=0$. This \ac{SB} configuration has shown strong robustness for both denoising and dereverberation~\cite{jukic2024schr}. Consequently, we get $\alpha_t = 1$ and $\sigma_t^2 = c(k^{2t}-1) / 2 \log k$. Due to the optimal transport characteristics of the \ac{SB}, the mean exactly interpolates between the clean speech $\mathbf x_0$ at $t = 0$ and the noisy speech $y$ at $t = 1$.

A key advantage of the \ac{SB} compared to \ac{SGMSE} is that the neural network $F_\theta$ can be trained to directly predict the data $\mathbf x_0$~\cite{chen_2023_sb}. This is in contrast to \ac{SGMSE}, where the Gaussian noise $\mathbf z$ is predicted as shown in Eq.~\eqref{eq:loss_score}. Using the data prediction loss, it has been proposed to include a time-domain auxiliary loss term based on the $\ell_1$-norm~\cite{jukic2024schr}. Additionally, we propose incorporating a perceptual loss term such that
\begin{equation}
  \label{eq:data_prediction_loss}
 \mathcal{L}_\text{SB} = \lVert F_\theta(\mathbf x_t, \mathbf y, t) - \mathbf{x}_0 \rVert_2^2 + \alpha \lVert \underline{\hat{\mathbf x}}_\theta - \underline{\mathbf{x}}_0 \rVert_1 - \alpha_\text{P} \operatorname{PESQ}(\underline{\hat{\mathbf x}}_\theta, \underline{\mathbf{x}}_0)\,,
\end{equation}
where $\underline{\hat{\mathbf x}}_\theta = \operatorname{iSTFT}(F_\theta(\mathbf x_t, \mathbf y, t))$ and $\underline{\mathbf{x}}_0 = \operatorname{iSTFT}(\mathbf x_0)$ represent the corresponding
time-domain signals using the \ac{iSTFT}. Moreover, $\operatorname{PESQ}(\cdot,\cdot)$ denotes a differentiable version of the PESQ metric\footnote{\url{https://github.com/audiolabs/torch-pesq}}, and $\alpha$ and $\alpha_\text{P}$ are hyperparameters to weight the different loss terms. 

At inference, the reverse SDE in Eq.~\eqref{eq:sb_reverse_sde} can be solved with an \ac{ODE} sampler or an \ac{SDE} sampler~\cite{chen_2023_sb}. Here, we make use of the \ac{ODE} sampler because it has shown better performance for the speech enhancement task~\cite{jukic2024schr}. For a given discretization schedule $(t_N=1, t_{N-1}, \dots, t_0=0)$ with $N$ steps, the ODE sampler is recursively defined as
\begin{equation}
    \mathbf x_{t_{n-1}} = a_{n} \mathbf x_{t_{n}} + b_{n} F_\theta(\mathbf x_{t_{n}}, \mathbf y, t_n) + c_{n} \mathbf y, \quad \mathbf x_{t_{N}} = \mathbf y\,,
\end{equation}
\begin{align}
    &a_{n} = \frac{\alpha_{t_{n-1}} \sigma_{t_{n-1}} \bar{\sigma}_{t_{n-1}}}{\alpha_{t_{n}} \sigma_{t_{n}} \bar{\sigma}_{t_{n}}}, \\
    &b_{n} = \frac{\alpha_{t_{n-1}}}{\sigma_1^2} \left( \bar{\sigma}_{t_{n-1}}^2 - \frac{\bar{\sigma}_{t_{n}} \sigma_{t_{n-1}} \bar{\sigma}_{t_{n-1}}}{\sigma_{t_{n}}} \right), \\
    &c_n = \frac{\alpha_{t_{n-1}}}{\alpha_1 \sigma_1^2} \left( \sigma_{t_{n-1}}^2 - \frac{\sigma_{t_{n}} \sigma_{t_{n-1}} \bar{\sigma}_{t_{n-1}}}{\bar{\sigma}_{t_{n}}} \right). 
\end{align}

\section{Experimental Setup}
\label{sec:experiments}

\begin{table}[t]
\centering
\begin{tabular}{lcrr}
\toprule
Model & \#$\,$params & GMACs & proc/s [s]\\ \midrule
Conv-TasNet~\cite{luo2019conv} &  8.7$\,$M & 28 & 0.015\\
MetricGAN+~\cite{fu2021metricganplus} & 1.9$\,$M & 106 & 0.016 \\ 
SGMSE+ \cite{richter2023speech} &  65.6$\,$M & 15,995 & 1.155 \\ 
SEMamba~\cite{chao2024investigation} &  2.3$\,$M & 131 & 0.075 \\ \bottomrule
\end{tabular}
\caption{Baseline methods. Number of parameters, GMACs for an input of 4 seconds, and average processing time.}
\label{tab:num_params_macs}
\end{table}

\subsection{Models settings}

We train eight models (M1-M8), each utilizing complex spectrograms as the input representation by computing the \ac{STFT} with a periodic Hann window of size of 510 and a hop length of 128. We use the identical amplitude compression as in \cite{richter2023speech}. We train the models with a batch size of 16 using two NVIDIA RTX A6000 \acp{GPU} with 48$\,$GB memory. 

Models M1-M4 employ the \ac{OUVE} \ac{SDE}, utilizing the recommended hyperparameters from \cite{richter2023speech}. The models vary in the loss type and the preconditioning, as shown in Table~\ref{tab:results}. 

Models M5-M8 use the \ac{SB-VE} with the recommended hyperparameters from \cite{jukic2024schr}. The models differ in the hyperparameter $\alpha_\text{P}$, as indicated in Table~\ref{tab:results}.

\subsection{Network architecture}

In all experiments except for model M4, we employ the NCSN++ architecture \cite{song2021sde} using the same parameterization described in \cite{richter2023speech}. 

For M4, we employ the EDM2 network architecture~\cite{karras2024analyzing}. The core idea in EDM2 is to restructure the network layers to ensure that the expected magnitudes of activations, weights, and updates maintain unit variance. Additionally, all additive biases are removed, and an extra channel of constant one is concatenated to the network's input instead. We use the same number of layers and channels as for the NCSN++.
Furthermore, the authors propose to use a power function \ac{EMA} that automatically scales according to training time and has zero contribution at the initial training step. 

\begin{table*}[t]
\centering
\begin{tabular}{@{}l|cccc|ccccc@{}}
\toprule
\textbf{Model} & \textbf{SDE} & \textbf{Loss} & $\alpha_\text{P}$ & \textbf{Precon} & \textbf{POLQA} & \textbf{PESQ} & \textbf{SI-SDR} & \textbf{ESTOI} & \textbf{DNSMOS} \\
\midrule
Noisy &   &   &   &   & $3.11 \pm 0.79$ & $1.97 \pm 0.75$ & $8.4 \pm 5.6$ & $0.79 \pm 0.15$ & $3.09 \pm 0.39$ \\
\midrule
Conv-TasNet+ \cite{luo2019conv} & - & - & - & - & $3.56 \pm 0.57$ & $2.63 \pm 0.60$ & $19.1 \pm 3.5$ & $0.85 \pm 0.10$ & $3.37 \pm 0.32$ \\
MetricGAN+ \cite{fu2021metricganplus} & - & - & - & - & $3.72 \pm 0.68$ & $3.13 \pm 0.55$ & $8.5 \pm 3.8$ & $0.83 \pm 0.11$ & $3.37 \pm 0.30$ \\
SEMamba \cite{chao2024investigation} & - & - & - & - & $\mathbf{4.33 \pm 0.40}$ & $3.56 \pm 0.60$ & $\mathbf{19.7 \pm 3.2}$ & $\mathbf{0.89 \pm 0.08}$ & $3.58 \pm 0.29$ \\
PESQetarian \cite{deoliveira2024pesqetarian} & - & - & - & - & $1.46 \pm 0.48$ & $\mathbf{3.82 \pm 0.57}$ & $-19.8 \pm 3.3$ & $0.84 \pm 0.09$ & $2.39 \pm 0.22$ \\
SGMSE+ \cite{richter2023speech} & OUVE & score & - & \xmark & $3.95 \pm 0.52$ & $2.93 \pm 0.62$ & $17.3 \pm 3.3$ & $0.87 \pm 0.10$ & $3.56 \pm 0.28$ \\
\midrule
M1 & OUVE & score & - & \xmark & $3.93 \pm 0.51$ & $2.84 \pm 0.61$ & $17.7 \pm 3.6$ & $0.86 \pm 0.10$ & $3.54 \pm 0.28$ \\
M2 & OUVE & denoise & - & \xmark & $3.96 \pm 0.53$ & $2.90 \pm 0.67$ & $18.0 \pm 3.3$ & $0.86 \pm 0.10$ & $3.55 \pm 0.28$ \\
M3 & OUVE & denoise & - & \cmark & $3.86 \pm 0.50$ & $2.77 \pm 0.59$ & $17.8 \pm 3.2$ & $0.86 \pm 0.10$ & $3.51 \pm 0.27$ \\
M4 (EDM2) & OUVE & denoise & - & \cmark & $3.87 \pm 0.54$ & $2.87 \pm 0.65$ & $18.0 \pm 3.2$ & $0.86 \pm 0.10$ & $3.54 \pm 0.27$ \\
M5 & SB-VE & predict & 0 & \xmark & $4.15 \pm 0.54$ & $2.91 \pm 0.76$ & $19.4 \pm 3.5$ & $0.88 \pm 0.10$ & $\mathbf{3.59 \pm 0.30}$ \\
M6 & SB-VE & predict & 1e-3 & \xmark & $4.15 \pm 0.53$ & $3.70 \pm 0.58$ & $8.3 \pm 2.8$ & $0.86 \pm 0.09$ & $3.44 \pm 0.34$ \\
M7 & SB-VE & predict & 5e-4 & \xmark & $4.25 \pm 0.50$ & $3.50 \pm 0.66$ & $14.1 \pm 2.9$ & $0.87 \pm 0.09$ & $3.55 \pm 0.29$ \\
M8 & SB-VE & predict & 2.5e-4 & \xmark & $4.20 \pm 0.51$ & $3.44 \pm 0.73$ & $15.3 \pm 2.8$ & $0.87 \pm 0.09$ & $3.58 \pm 0.29$ \\
\bottomrule
\end{tabular}
\caption{Speech enhancement performance on VB-DMD. Values indicate mean and standard deviation.}
\label{tab:results}
\vspace{-1em}
\end{table*}

\subsection{Metrics}

As intrusive speech enhancement metrics, we include POL\-QA~\cite{polqa2018} and PESQ~\cite{rixPerceptualEvaluationSpeech2001} for predicting speech quality. Moreover, we employ ESTOI~\cite{jensen2016algorithm} as an instrumental measure of speech intelligibility and calculate the \ac{SI-SDR}~\cite{leroux2018sdr} measured in dB. As a non-intrusive metric, we use DNSMOS~\cite{reddy2021dnsmos}, which employs a neural network trained on human ratings. For all metrics it holds, the higher, the better.

\subsection{Baselines and Data}

Table~\ref{tab:num_params_macs} shows all baseline methods, the number of parameters, \acfp{MAC} for an input of 4 seconds, and the processing time per input second on a \ac{GPU}. We use the provided checkpoints and the official implementations. As a dataset, we use the standardized \ac{VB-DMD}~\cite{valentini2016investigating}, commonly employed as a benchmark for speech enhancement.

\section{Results}
\label{sec:results}

\begin{figure}[t]
    \centering
    \includegraphics[scale=0.7]{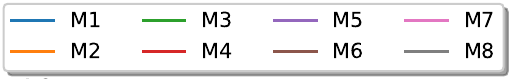}
    \\
    \vspace{0.3em}
    \begin{subfigure}{0.49\linewidth}
        \centering
        \includegraphics[width=\linewidth]{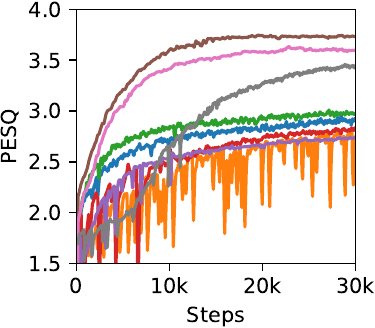}
        \label{fig:pesq}
    \end{subfigure}
    \hfill
    \begin{subfigure}{0.49\linewidth}
        \centering
        \includegraphics[width=\linewidth]{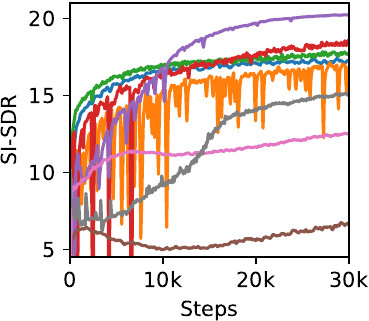}
        \label{fig:si_sdr}
    \end{subfigure}
        \vspace{-2em}
    \caption{PESQ and SI-SDR performance over the training steps.}
    \label{fig:training}
\end{figure}

In Table \ref{tab:results}, we present the results for the speech enhancement task using the \ac{VB-DMD} dataset. We begin by comparing different training objectives for the \ac{OUVE} \ac{SDE}. Model M1 employs the standard \ac{SGMSE} training objective, replicating the results reported in the original paper \cite{richter2023speech}. Minor discrepancies in these outcomes may be attributed to a different batch size. Model M2 utilizes the denoising loss in Eq.~\eqref{eq:denoising_loss} and achieves slightly higher scores than M1. However, its training process is more unstable, as depicted by the orange lines in Fig. \ref{fig:training}. In contrast, the preconditioned model M3 displays faster training times but produces results comparable to those of M1. Model M4 explores the EDM2 network, exhibiting performance that is competitive with M1. Yet, similar to M2, it experiences instability during training, indicated by the red lines in Fig. \ref{fig:training}. It is worth noting that the results for M4 are preliminary, as we have not experimented with various post-hoc \ac{EMA} configurations and fine-tuned learning rate schedulers. 
Among models M1-M4, differences in ESTOI and DNSMOS are negligible.

Turning the attention to the \ac{SB} approach, model M5 shows strong performance in \ac{SI-SDR}; an expected result due to the $\ell_1$-loss applied in the time domain. Moreover, M5 demonstrates relatively stable training behavior, as illustrated by the purple lines in Fig.~\ref{fig:training}, and achieves the best score for DNSMOS among all methods.  Model M6 introduces the PESQ loss and achieves state-of-the-art results in PESQ with a score of 3.70. However, similar to the extreme behavior of the PESQetarian~\cite{deoliveira2024pesqetarian}, this improvement comes at the expense of SI-SDR. Models M7 and M8 investigate different weightings of the PESQ term, achieving a better balance between SI-SDR and PESQ, while also obtaining higher ESTOI scores compared to M5 and M6. Specifically, M7 performs best in POLQA among all models based on the \ac{SB}.

Lastly, we examine a comparison with the given baseline methods. With the \ac{SB} approach, we surpass the performance of most baselines, including \ac{SGMSE} while remaining on par with SEMamba~\cite{chao2024investigation}. 
For results obtained with 48$\,$kHz speech data using the EARS-WHAM dataset~\cite{richter2024ears}, we refer  to~\cite{richter2024diffusion}. Audio files are available online\footnote{\url{https://sp-uhh.github.io/gen-se/}}

\section{Conclusion}
\label{sec:conclusion}

This paper explored the distinctions among various dif\-fu\-sion-based frameworks for generative speech enhancement, explicitly focusing on score-based generative models and the Schrödinger bridge (SB). Through comprehensive experimental analysis, we highlighted the variations in training behaviors and performance across these frameworks. We proposed adding ad-hoc cost functions to the \ac{SB} framework, significantly enhancing the performance and perceptual quality of the processed speech signals. To support ongoing research and innovation in this field, we made all experimental code and pre-trained models publicly accessible.

\bibliographystyle{IEEEtran}
\bibliography{refs}

\end{document}